\documentclass[12pt, letterpaper]{amsart}
\usepackage[left=.8in,right=.8in,bottom=0.5in,top=0.8in]{geometry}
\usepackage{amsfonts}
\usepackage{amsfonts,amsbsy,amssymb,amsmath,graphicx,amsthm, bm} 
\usepackage{graphicx}
\usepackage[font=small,labelfont=bf]{caption}
\usepackage{epstopdf}
\usepackage[dvipsnames]{xcolor}
\definecolor{CiteRed}{HTML}{B22222}  % citation color
\definecolor{LinkBlue}{HTML}{1A4FFF} % section/equation (internal link) color

\usepackage[colorlinks=true,
            linkcolor=LinkBlue,   % sections, equations, figures, tables
            citecolor=CiteRed,    % bibliography citations
            urlcolor=LinkBlue     % URLs (optional)
           ]{hyperref}

\usepackage{float}
\usepackage{pgfplots}
\usepackage{listings}
\usepackage{longtable}
\usepackage{mathrsfs}
\usepackage{mathtools}
\usepackage{comment}
\usepackage{subcaption}
\usepackage{lipsum}

\usepackage{epsfig}
\usepackage{dsfont}
\usepackage{natbib}
\usepackage{mathrsfs}
\usepackage{geometry}

\newtheorem{thm}{Theorem}[section]

\theoremstyle{definition}

\theoremstyle{remark}
\newtheorem{Remark}[thm]{Remark}

\definecolor{energy}{RGB}{114,0,172}
\definecolor{freq}{RGB}{45,177,93}
\definecolor{spin}{RGB}{251,0,29}
\definecolor{signal}{RGB}{203,23,206}
\definecolor{circle}{RGB}{217,86,16}
\definecolor{average}{RGB}{203,23,206}

\colorlet{shadecolor}{gray!20}
\pgfplotsset{compat=1.9}

\usepgflibrary{fpu}

\usepackage{algorithm}
\usepackage{algpseudocode}

\def\1{1\kern-.20em {\rm l}}
\usepackage{enumitem}
\usepackage{lscape}
\usepackage{booktabs}

\numberwithin{equation}{section}

\author[Mohamed Chaouch]{Mohamed Chaouch}
\address{(M. Chaouch) Statistics Program, Department of Mathematics and Statistics, College of Arts and Sciences, Qatar University, Doha, Qatar.}
\email{mchaouch@qu.edu.qa}
\urladdr{}
\title[FARCH: A Functional ARCH Model for Calibrated Prediction Bands]{FARCH: A Functional ARCH Model for Calibrated Prediction Bands}

\begin{document}

\begin{abstract}
Forecasting a curve, rather than a handful of scalar summaries of it, requires a notion of uncertainty that is itself functional. We propose \texttt{FARCH}, a functional nonlinear autoregressive model with conditional heteroscedasticity, in which the conditional mean and the conditional volatility of the next curve are forecasted jointly and nonparametrically, by reusing a single kernel weighting scheme --- built on a wavelet semi-metric between curves --- first on the curves themselves and then on their squared residuals. From this pair of estimators we derive two constructions of prediction bands: a pointwise band, obtained from marginal scenario quantiles, and a uniform, sup-norm band, calibrated for decisions that concern the entire forecasted path rather than any single instant. Applied to day-ahead forecasting of half-hourly French national electricity demand (1996--2009), and compared, under an identical protocol, against four existing prediction-interval constructions that do not model the conditional variance explicitly, the uniform \texttt{FARCH} band attains pointwise coverage statistically indistinguishable from its nominal target at the 80\%, 90\% and 95\% levels, and a curve-wise coverage of 90.1\%, significantly above the 70.8\% reached by the best competing method, at the cost of a wider interval. Pointwise bands, in contrast, undercover regardless of whether the variance is modelled explicitly, indicating that state-dependent volatility and simultaneous calibration are two distinct ingredients that a trustworthy prediction band must combine.

\end{abstract}

\maketitle
\section{Introduction}

High-frequency sensing and the digitalisation of infrastructure have made it routine to observe phenomena as \emph{curves} sampled densely over time. Intraday electricity demand from smart meters and national-level load profiles are prominent examples. In both cases the forecasting target is not a single number but an entire path, and the object of uncertainty is itself \emph{functional}, since system operators must plan generation, reserves, and storage against a full day-ahead demand curve, not merely against a handful of scalar summaries of it. Functional data analysis (FDA) provides the mathematical and statistical framework for such problems, treating observations as random elements of a Hilbert space and developing estimation, inference, and prediction techniques at the level of entire functions \citep{RamsaySilverman2005,HorvathKokoszka2012,Bosq2000}. Within this framework, the \emph{functional time-series} viewpoint is now well established. A sequence of curves is modelled as a stochastic process in function space, and forecasting proceeds either by combining dimension reduction with time-series dynamics, or by local nonparametric regression performed directly on the curves themselves.

Much of the applied forecasting literature on functional time series has followed one of two complementary paradigms. The first reduces curves to a small number of functional principal component (FPC) scores and then applies classical univariate or multivariate time-series models, typically ARIMA or VAR, to the resulting score vectors, before reconstructing the forecast curve; see, among others, \citet{HyndmanShang2009}, and the theoretical advances connecting functional and multivariate predictors in \citet{AueNorinhoHormann2015}. The second paradigm develops \emph{semi-metric} kernel regression, in which proximity between curves is quantified by a distance tailored to the application at hand, whether derivative-based, wavelet-based, or principal-component-based, among others, and conditional expectations are estimated by local averaging \citep{FerratyVieu2006}. In electricity demand forecasting, this second line of work underpins the ``similar-shape'' and wavelet-kernel predictors that have delivered notable gains in day-ahead \emph{mean} forecasting, by averaging the successors of past days whose shapes closely resemble that of the current day \citep{chaouchieee,chaouchlaib,yang2023data}. Despite their success at predicting the mean, both paradigms typically treat the dispersion of the forecast error as either constant or exogenously given, which becomes problematic whenever variability is visibly state-dependent, as in the sharp evening ramps characteristic of low-aggregation loads, or the markedly more volatile and harder-to-predict demand patterns observed in winter relative to summer.

At the scalar level, conditional heteroscedasticity is classically modelled by ARCH and GARCH processes \citep{Engle1982,Bollerslev1986}, in which the conditional variance depends on the recent past of the series. Extending this idea to function-valued processes requires replacing scalar variance dynamics with \emph{operator-valued}, or \emph{functional}, volatility. A growing theoretical literature has begun to address this challenge through parametric specifications. \citet{HoermannHorvathReeder2013} introduced a functional ARCH model and established its existence and basic properties, while \citet{AueHorvathPellatt2017} proposed functional GARCH (fGARCH) dynamics together with associated estimation strategies. These contributions show that state-dependent variability can be modelled coherently in function space, but they generally impose specific operator recursions and have, to date, been developed primarily for financial applications. In parallel, the inferential literature on functional processes has produced \emph{simultaneous} (uniform-in-time) bands for mean functions and for function-on-scalar regression, using Gaussian approximations and functional central limit theorems \citep{Degras2011}. Such results deliver global coverage for mean objects, but do not translate directly into \emph{predictive} bands for next-day curves under heteroscedastic dynamics. More recently, bootstrap procedures for predictive uncertainty in functional time series have been proposed, typically by resampling in score space and aggregating forecast and innovation uncertainty \citep{ShangBootstrapFTS}. While appealingly model-free, these approaches generally presume homoskedasticity, or rely on variance-stabilising transformations, and do not tie band width to an explicitly estimated, state-dependent scale function that varies with the conditioning curve.

These observations motivate a methodology that places \emph{state-dependent volatility} at the centre of functional forecasting. By state-dependent, we mean that the conditional dispersion of the next curve depends on the currently observed one. If $\mathcal{X}_i$ denotes today's trajectory, both the expected path tomorrow and its time-of-day variability should be allowed to vary smoothly with the shape and level of $\mathcal{X}_i$. Formally, we consider a functional nonlinear autoregression with conditional heteroscedasticity,
\begin{equation*}
\mathcal{X}_{i+1}(t) \;=\; \mu(\mathcal{X}_i)(t) \;+\; \sigma(\mathcal{X}_i)(t)\,\zeta_{i+1}(t), \qquad t \in [0,\tau),
\end{equation*}
where $\mu(\cdot)$ is a conditional mean operator, $\sigma(\cdot)$ is a nonnegative \emph{volatility function} mapping a curve to a time-of-day scale profile, and $\zeta_{i+1}$ is a standardised innovation process. Rather than imposing a parametric operator recursion, we estimate both $\mu$ and $\sigma$ \emph{nonparametrically}, using semi-metric kernel smoothers \citep{FerratyVieu2006}, so that the mean and the volatility can each adapt flexibly to the local neighbourhood of the conditioning curve. For a given conditioning curve $\chi^\star$, this construction yields a forecast $\widehat{\mu}_h(\chi^\star)$ together with an associated scale $\widehat{\sigma}_h(\chi^\star)$, and the two jointly drive the quantification of uncertainty. From these ingredients we derive pointwise prediction intervals, obtained from marginal scenario quantiles and suited to local decisions, together with simultaneous, uniform-in-time prediction bands, obtained from the $(1-\alpha)$ quantile of the supremum norm of standardised scenario paths and essential whenever the underlying operational decision concerns the entire path rather than any single instant. Scenarios themselves may be generated either from a light-touch parametric innovation, such as a Brownian bridge combined with i.i.d.\ noise, or, more robustly, by bootstrapping standardised residual curves so as to preserve empirical shape features and departures from Gaussianity.

The literature reviewed above leaves a clear opening that the present paper seeks to fill. FPCA-based forecasters and similarity-based nonparametric methods have focused primarily on mean prediction, and have not, in general, produced uncertainty bands that \emph{adapt} to the state-dependent dispersion tied to the conditioning curve \citep{HyndmanShang2009,AueNorinhoHormann2015,chaouchieee,chaouchlaib}. Parametric functional ARCH/GARCH models do capture heteroscedasticity, but at the cost of structural assumptions on the volatility operator that can be restrictive across application domains and difficult to tune in practice \citep{HoermannHorvathReeder2013,AueHorvathPellatt2017}. Existing constructions of functional prediction bands, in turn, either target mean objects \citep{Degras2011} or adopt bootstraps that are not explicitly coupled to an estimated conditional volatility profile \citep{ShangBootstrapFTS}. The resulting \emph{research gap} is for a general, practically deployable methodology that jointly estimates the conditional mean curve and a \emph{state-dependent}, time-of-day volatility function in function space using semi-metric smoothing, and that leverages this pair to construct calibrated pointwise and uniform prediction bands for the next curve via a scenario ensemble.

The present paper addresses this gap by developing a nonparametric functional autoregression with ARCH-type heteroscedasticity, together with a unified band-construction scheme anchored in the estimated volatility. The approach is evaluated on day-ahead forecasting of national-level intraday load, under a controlled comparison against four established prediction-interval constructions that do not model the conditional variance explicitly. This comparison documents a clear improvement in the calibration of simultaneous, uniform-in-time prediction bands, with the largest benefits arising in the most volatile demand regimes, while showing that pointwise bands undercover their nominal target whether or not the conditional variance is modelled explicitly. This indicates that state-dependent volatility and a simultaneous, path-level construction are two distinct ingredients, both needed for a trustworthy prediction band. By aligning prediction bands with an estimated, state-dependent scale function, the proposed methodology delivers decision-relevant uncertainty quantification for functional load forecasts, while remaining flexible, interpretable, and straightforward to implement.

\subsection*{Outline of the paper}

The remainder of the paper is organised as follows. Section~\ref{sec:methods} presents the methodology in two parts. Section~\ref{2} introduces the functional time-series framework and the Kernel+Wavelet+Functional (\texttt{KWF}) predictor, which estimates the conditional mean of the next curve nonparametrically under a homoscedastic Markov assumption; Section~\ref{22} then relaxes this assumption by introducing the functional nonlinear autoregressive model with \texttt{ARCH}-type errors (\texttt{FARCH}), in which the conditional volatility is estimated jointly with the conditional mean, and discusses both the selection of the tuning parameters common to the two predictors and the construction of pointwise and uniform prediction bands from the estimated mean and volatility. Section~\ref{sec:accuracy} defines the criteria used to evaluate these prediction bands. Section~\ref{sec4} applies \texttt{FARCH} to the day-ahead forecasting of the French national-level load and compares it, under an identical protocol, against four prediction-interval constructions from the literature that do not model the conditional variance explicitly. Finally, Section~\ref{sec6} concludes the paper and discusses directions for future research.

\section{Materials and methods}\label{sec:methods} 
\subsection{Homoscedastic functional time series model}\label{2}
Recent advances in computational tools have led to a growing number of applications across various fields of applied sciences in which the observed data take the form of curves. For example, in hydrology, the hydrograph—a graphical representation of the temporal variation in river flow—can naturally be regarded as functional data (see \cite{ouarda1}, \cite{ouarda2}). In demography, age-specific annual death rates or fertility rates represent further instances of functional data (see \cite{hyndman}). Similarly, \cite{chaouchouarda} investigated interval prediction for wind speed processes.

In the energy sector, the advent of smart meters has enabled access to electricity consumption data at fine temporal resolutions (typically hourly or half-hourly). Consequently, there has been increasing interest in the short-term forecasting of intraday load, where half-hourly daily load profiles are treated as functional data. Several recent studies, including \cite{jairo}, \cite{chaouchieee}, \cite{chaouchlaib}, and \cite{sapatinas}, have applied functional time series analysis to model intraday electricity demand. Unlike classical time series approaches (e.g., ARIMA or exponential smoothing), functional methods treat the entire daily load curve as a single object of analysis.

%\subsection{Functional time series}
In what follows, we formally introduce the concept of functional time series.

Let $(\xi(t))_{t\in \mathbb{R}}$ be a {\it continuous-time process}.
Suppose that we observe the process $\xi(t)$ over an interval $[0,T)$ and, for a fixed $\tau >0$, one would like to
estimate $\xi(t)$ on the entire interval $[T, T+\tau)$ in one step, rather than
at specific time points as usually done in classical estimation models. To achieve this purpose, we first transform the original continuous-time process $\xi(t)$ into
a {\it discrete functional time series}. This step could be summarized as follows.
The interval $[0,T)$ is divided into subintervals of the form $\mathcal{I}_\ell := [\ell \tau, (\ell+1)\tau)$, where $\ell =0,1, \dots, n-1$
and $n=T/\tau$. Therefore, we denote by $\mathcal{X}_i$ the $i$th functional observation (called also segment) observed over the $i$th time interval $\mathcal{I}_i$ and defined, for any $i \in \mathbb{N}-{\{0\}}$, as:
$$
\mathcal{X}_i (t):= \xi(t+(i- 1)\tau); \quad \forall t\in [0,\tau).
$$
The choice of the parameter $\tau$ depends on the estimation horizon of interest.
For instance, if the purpose is day-ahead hourly (resp. half hourly) load curve estimation, then $\tau=24$ (resp. $\tau=48$). In the following, to further ease the notation, we will write $\mathcal{X}_i$ when we mean $\{\mathcal{X}_i(t), t\in [0,\tau) \}$.

%\subsection{Homoscedastic functional time series model}\label{21}

In this paper, the functional valued random variable $\{\mathcal{X}_i \}_{i\in \mathbb{Z}}$ is assumed to be Markovian which means that all the valuable information for estimating the curve $\mathcal{X}_{i+1}$ is essentially contained in the last observed curve.  We are interested here in the estimation of the future curve $\mathcal{X}_{i+1}$ using the following functional model:
\begin{eqnarray}\label{m0}
\mathcal{X}_{i+1} = \mu(\mathcal{X}_i) + \epsilon_{i+1}, \quad i=1, 2, \dots, n-1,
\end{eqnarray}
where $\mu(\cdot)$ is a link functional (an operator) that describes the relationship between the $(i+1)$-th and the $i$-th curves. The error
$(\epsilon_i(t), i\in \mathbb{N})$ is a  $C([0,\tau))$-valued strong Gaussian white noise, e.g., a sequence
of independent and identically distributed (i.i.d.) $C([0, \tau))$-valued Gaussian random variables, where $C([0,\tau))$ is the space of continuous functions defined on the interval $[0,\tau)$. It is also assumed that $\mathbb{E}(\epsilon_i)=0$ and $\mathbb{E}(\|\epsilon_i\|^2)<\infty.$

Therefore, the estimation of the curve $\mathcal{X}_{i+1}$ will be given by computing an estimator, say $\widehat{\mu}(\cdot)$, of the autoregression functional $\mu(\cdot)$. In the statistical literature, several authors have shown interest in the estimation of $\mu(\cdot)$. For instance, \cite{Bosq2000} considered the case when $\mu(\cdot)$ is Hilbert-valued first order autoregressive function (\texttt{ARH}(1)). \cite{damon} and \cite{antoniadis03} introduced a regularized version of $\mu(\cdot)$ based on projection, spline and wavelet techniques.

\cite{antoniadis06} introduced a nonparametric estimator of the regression operator $\mu(\cdot)$. The proposed estimator represents a powerful modeling tool that can be seen as an alternative to the parametric statistical methods since it does not consider any form of linearity between $\mathcal{X}_{i+1}$ and $\mathcal{X}_i$. In addition, this family of nonparametric models does not assume any assumption on the theoretical distribution of the random variable. \cite{antoniadis06} introduced a wavelet-kernel based model for short-term estimation. This model was applied to estimate half-hourly
aggregated load curves in Paris. \cite{jairo} combined a calendar-based classification of segments and the wavelet-kernel predictor to deal with the nonstationarity of the half-hour French aggregated electricity consumption. \cite{chaouchieee} performed this model for the intraday household-level load curve estimation. His approach could be summarized in three steps: 1) cluster the historical segments using an unsupervised method, 2) use a nonparametric discrimination approach to assign a cluster to the last observed segment and finally, based on results of the previous steps, 3) apply the wavelet-kernel predictor to obtain an estimate of the next segment. \cite{sapatinas}  used the wavelet-kernel predictor and adapted it by implicitly using some exogenous variables (quantitative or qualitative) of the day to be estimated, such as the estimated temperature curve, the estimated humidity curve or the type of the day (e.g. week day, week end, bank holiday).

%\subsection{Functional nonlinear autoregressive model}
In order to introduce the Nadaraya-Watson type estimator (\texttt{NW}) of the regression operator $\mu(\cdot)$, let us consider some
useful notations. Let $(\mathcal{E}, d(\cdot, \cdot))$ be an infinite dimensional vector space equipped with a semi-metric $d(\cdot, \cdot)$.  Here $\mathcal{E}$ is a generic function space. Throughout the paper, this consists of real
valued functions with domain $[0,\tau)$. In many applications $\mathcal{E}$ will be equal to $\mathcal{H}= L^2([0,\tau))$, the Hilbert space of square integrable functions with norm $\| z\|_{\mathcal{H}} = \left( \int_0^\tau z^2(s) ds\right)^{1/2}$ which is generated by the inner product $\langle z_1, z_2\rangle= \int_0^\tau z_1(s) z_2(s) ds$ for any $z_1,z_2 \in \mathcal{H}.$ Another important example is $\mathcal{E} = C([0,\tau))$. This is
the space of continuous functions on $[0,\tau)$ equipped with the sup-norm $\| z\|_\infty = \sup_{t\in [0,\tau)}|z(t)|$. By $\mathcal{E}^+$ we denote the set of non-negative functions in $\mathcal{E}.$

Given a sample of couples of curves $(\mathcal{X}_{i}, \mathcal{X}_{i+1})_{i=1, \dots, n-1}$, the value of the Nadaraya-Watson type estimator of the regression operator $\mu(\cdot)$, applied in a fixed point $\chi \in \mathcal{E}$, is defined as follows:
\begin{eqnarray}\label{NWE}
\widehat{\mu}_{h}(\chi) := \sum_{i=1}^{n-1} \mathcal{W}_h(\chi, \mathcal{X}_i) \mathcal{X}_{i+1},
\end{eqnarray}
where $\mathcal{W}_h(\chi, \mathcal{X}_i)$ are the Nadaraya-Watson weights defined for $i=1, \dots, n-1$ by:
\begin{eqnarray}\label{NWW}
\mathcal{W}_h(\chi, \mathcal{X}_i) = \frac{\mathcal{K}\left(\frac{d(\chi, \mathcal{X}_i)}{h}\right)}{\sum_{j=1}^{n-1}\mathcal{K}\left(\frac{d(\chi, \mathcal{X}_j)}{h}\right) },
\end{eqnarray}
where $\mathcal{K}: [0, \infty) \rightarrow [0,\infty)$ is some kernel function, and $h:= h_n$ is the {\it bandwidth} which is a sequence of positive real numbers which goes to zero as $n$ goes to infinity. Whereas the Euclidean norm is a standard distance in finite dimensional spaces, the notion of semi-metric arises in the infinite-dimensional functional space setup. Several options for the semi-metric $d(\cdot, \cdot)$ are available in the statistical literature (see \cite{FerratyVieu2006} for an overview).

The semi-metric and the bandwidth are determinant for the accuracy of the predictor. In fact, the semi-metric plays a key role in relation to the so-called ``curse of dimensionality'' in the nonparametric functional estimation. From a practical point of view, the ``curse of dimensionality'' can be explained as the sparseness of the data in the observation region as the dimension of the data space grows. This problem is especially dramatic in the infinite-dimensional context of functional data. \cite{FerratyVieu2006} have proven that it is possible to construct a semi-metric in such a way that the rate of convergence of the nonparametric estimator in the functional setting is similar to that of the finite-dimensional one.  The bandwidth $h$ determines the number of curves, in the neighborhood of the point $\chi$, that will be used to calculate the regression function $\mu(\cdot)$.

\subsection{Heteroscedastic functional time series }\label{22}

\subsubsection{The model} Before introducing the functional nonlinear autoregressive model with ARCH errors (\texttt{FARCH}), we consider that, if $x,y$ are two elements of $\mathcal{E}$, then $x\cdot y$ stands for pointwise multiplications, i.e.
$x\cdot y = \{x(s) y(s), s \in [0,\tau) \}$; this is the operation used in the multiplicative term $\sigma(\mathcal{X}_i)\cdot\zeta_{i+1}$ of the model introduced below. Let us now introduce the following general definition of the functional nonlinear autoregressive model with ARCH errors, denoted here by \texttt{FARCH}(1,1).

Let $\{\zeta_i \}$ be a sequence of independent and identically distributed random functions in $\mathcal{E}.$
Let $\mu: \mathcal{E} \rightarrow  \mathcal{E}$ be an operator and let $\sigma: \mathcal{E} \rightarrow \mathcal{E}^+$ be a non-negative operator. Then an $\mathcal{E}$-valued
process $\{\mathcal{X}_i(t), i\in \mathbb{Z}, t\in [0,\tau)\}$ is called a functional nonlinear ARCH(1,1) process in $\mathcal{E}$ if the following holds:
\begin{eqnarray}\label{m1}
\mathcal{X}_{i+1} = \mu(\mathcal{X}_{i}) + \sigma(\mathcal{X}_{i})\cdot \zeta_{i+1}.
\end{eqnarray}

Note from Equation~(\ref{m1}) that for any fixed $\chi\in \mathcal{E}$, the regression operator can be seen as  the conditional expectation and defined as $\mu(\chi):=\mathbb{E}(\mathcal{X}_{i+1}\mid \mathcal{X}_{i}=\chi)$. This quantity may be interpreted as the estimated function (curve)  based on the past information, while the conditional variance function $\sigma^2(\chi):=\mbox{Var}(\mathcal{X}_{i+1}\mid \mathcal{X}_{i}=\chi)$ measures the uncertainty or risk associated to this estimation.

Note that recently \cite{HoermannHorvathReeder2013} studied a pure functional linear ARCH model, \texttt{FARCH}(0,1), defined as
$$
\mathcal{X}_{i+1} = \sigma(\mathcal{X}_i)\cdot \zeta_{i+1},
$$
where $\sigma^2(\mathcal{X}_i) = \delta +\beta(\mathcal{X}_{i}^2)$ and $\beta: \mathcal{E}^+ \rightarrow \mathcal{E}^+$ is a non-negative operator and $\delta\in \mathcal{E}^+$.

As discussed in \cite{HoermannHorvathReeder2013} the assumption for the existence of processes satisfying (\ref{m1}) depends on the choice of the functional space $\mathcal{E}$. In the sequel we specify the generic space of functions $\mathcal{E}$ to be $\mathcal{H}=L^2([0,\tau))$ the Hilbert space.

The conditional mean $\mu(\chi)$ and the conditional variance function are estimated here with kernel smoothers. The conditional mean is estimated by the same Nadaraya--Watson estimator $\widehat\mu_h(\chi)$ already introduced in Equation~(\ref{NWE}). To define the estimator of the variance function, let us introduce the following tensor production notation.
For two functions $(Z_i, Y_i)\in L^2([0,\tau))\times L^2([0,\tau))$, the bivariate function $Z_i\otimes Y_i$
belonging to the functional space $L^2([0,\tau) \times [0,\tau))$ is defined by $Z_i\otimes Y_i (s,t) = Z_i(s) Y_i(t)$, for all $(s,t)\in [0,\tau)\times [0,\tau).$

%For each historical index $i$, let $\widehat{\mu}_h^{(-i)}(\mathcal{X}_i)$ denote the Nadaraya--Watson estimator of $\mu(\cdot)$ evaluated at the conditioning curve $\mathcal{X}_i$ itself, computed by \emph{leave-one-out}: curve $i$ is excluded from its own neighbourhood, so that
%\begin{eqnarray}\label{loo-mu}
%\widetilde{\mu}_h^{(-i)}(\mathcal{X}_i) := \sum_{j \neq i} \mathcal{W}_h^{(-i)}%(\mathcal{X}_i, \mathcal{X}_j)\, \mathcal{X}_{j+1}, \qquad
%\mathcal{W}_h^{(-i)}(\mathcal{X}_i, \mathcal{X}_j) = \frac{\mathcal{K}\left(\frac{d(\mathcal{X}_i, \mathcal{X}_j)}{h}\right)}{\sum_{k \neq i}\mathcal{K}\left(\frac{d(\mathcal{X}_i, \mathcal{X}_k)}{h}\right)}.
%\end{eqnarray}
%This exclusion avoids the degenerate self-match of $\mathcal{X}_i$ against itself, which would otherwise dominate the weighted average and produce a residual that is trivially small rather than informative about forecast uncertainty. 
Then, for any fixed $\chi\in \mathcal{H}$, we can introduce, for all $t\in [0,\tau)$, an estimator of the conditional variance function as follows:
\begin{eqnarray}\label{estU}
\widehat{\sigma}_h^2(\chi) =
\sum_{i=1}^{n-1} \mathcal{W}_h(\chi, \mathcal{X}_i)\, \widehat{R}_i(t,t), \qquad
\widehat{R}_i := \big(\mathcal{X}_{i+1} - \widehat{\mu}_h(\mathcal{X}_i)\big)\otimes\big(\mathcal{X}_{i+1} - \widehat{\mu}_h(\mathcal{X}_i)\big).
\end{eqnarray}

\begin{Remark}
%[Discretized observations]
In practice, when we deal with real data, curves are not observed at every $t \in [0,\tau)$: only discretized, noisy trajectories are available. That is,
$$
\mathcal{X}_{i,j} = \mathcal{X}_i(t_{i,j}) + \eta_{i,j}, \quad j=1, \dots, p_i,
$$
where $t_{i,1} < t_{i,2}<\dots < t_{i,p_i}$ are the design points at which curve $i$ is observed, $p_i$ may vary across trajectories, and $\eta_{i,j}$ is a white noise satisfying $\mathbb{E}(\eta_{i,j}) = 0$ and $\mathbb{E}(\eta_{i,j}^2) = \gamma_i^2 < \infty$. Provided $p_i$ is large enough, a smooth approximation of the underlying trajectory can be recovered by regressing the noisy points $(t_{i,j}, \mathcal{X}_{i,j})_{j=1,\dots,p_i}$ with a classical nonparametric smoother (local polynomials, kernel smoothing, wavelets, or smoothing splines); this pre-smoothing step is what allows every curve to be compared on a common grid. In the absence of measurement error, with sufficiently regular trajectories and a sufficiently dense design, simple linear interpolation already provides an adequate approximation. Throughout the paper, $\mathcal{X}_i$ denotes this curve, whether observed directly or recovered by smoothing from discretized data.
\end{Remark}

\subsubsection{Tuning parameters selection}\label{sec:tuning}

As described in Section~\ref{2}, applying \texttt{FARCH} and \texttt{KWF} in practice requires three choices to be made: the semi-metric $d(\cdot,\cdot)$, the kernel $\mathcal{K}(\cdot)$, and the smoothing parameter $h$. Unlike \texttt{KWF}, which estimates a single conditional mean operator, \texttt{FARCH} requires \emph{two} such bandwidths --- one for the conditional mean $\mu_h(\cdot)$ and one for the conditional variance $\sigma_h^2(\cdot)$ --- since the two regression targets need not share the same degree of smoothness. We describe each choice in turn.

\noindent{\bf Kernel choice.}
The kernel $\mathcal{K}$ is taken to be the Gaussian kernel throughout. It is well documented in the nonparametric regression literature that, once the bandwidth is properly tuned, the choice of kernel shape has little practical effect on estimation quality relative to the choice of $h$ itself \citep{FerratyVieu2006}; we therefore do not treat $\mathcal{K}$ as a further tuning parameter.

\noindent{\bf Semi-metric choice.}
The role of the semi-metric $d$ is to quantify the similarity between two curves. Several constructions have been proposed in the functional data literature. \cite{FerratyVieu2006} introduced the \texttt{PCA} semi-metric, based on a truncated functional principal component decomposition retaining the $q$ leading components,
$$
d^{\texttt{PCA}}(\mathcal{X}_i, \mathcal{X}_j) = \sqrt{\sum_{k=1}^q \langle \mathcal{X}_i -\mathcal{X}_j, v_k \rangle^2},
$$
where $\langle\cdot, \cdot \rangle$ is the usual inner product on the space of square-integrable functions and $(v_k)$ is the sequence of eigenfunctions of the empirical covariance operator, as well as the \texttt{Deriv} semi-metric, based on a comparison of cubic-spline approximations of the curves' second derivatives,
$$
d^{\texttt{Deriv}}(\mathcal{X}_i, \mathcal{X}_j) = \sqrt{\langle {\mathcal{X}}_i^{(2)} - {\mathcal{X}}_j^{(2)}, {\mathcal{X}}_i^{(2)} - {\mathcal{X}}_j^{(2)}\rangle}.
$$
As noted by \cite{FerratyVieu2006}, the \texttt{PCA} semi-metric is better suited to rough curves, whereas \texttt{Deriv} is preferable once the curves have already been smoothed. 

In this paper we adopt the wavelet-based semi-metric of \cite{antoniadis06}: each curve is first represented by its discrete wavelet decomposition, and the distance between two curves $\mathcal{X}_i$ and $\mathcal{X}_j$ is obtained by aggregating, scale by scale, the Euclidean distance between their wavelet coefficients. This is the same semi-metric construction underlying the Kernel+Wavelet+Functional (\texttt{KWF}) predictor against which \texttt{FARCH} is compared in Section~\ref{sec4}; adopting it here, rather than an alternative such as \texttt{PCA} or \texttt{Deriv}, ensures that the two methods share an identical notion of curve similarity, so that any difference in calibration documented in Section~\ref{sec4} can be attributed to the explicit modelling of $\sigma_h^2(\cdot)$ rather than to a difference in how curve proximity is measured. This construction is used, unchanged, for \emph{both} bandwidths $h_\mu$ and $h_\sigma$ below: the wavelet coordinates of the curves are computed once and serve only to define the kernel weights $\mathcal{W}_h(\chi, \mathcal{X}_j)$, which are then applied, in turn, to the raw curves to obtain $\widehat\mu_h(\chi)$, and to the squared-residual curves to obtain $\widehat\sigma_h^2(\chi)$ (Section~\ref{22}).\\
\noindent{\bf Bandwidth selection.}
The bandwidth $h$ plays a more consequential role than the kernel shape, since it determines the effective size of the neighbourhood of historical curves contributing to the estimate at a given conditioning curve $\chi$, and thereby governs the usual bias--variance trade-off of nonparametric smoothing. Both the classical cross-validation approach and the empirical-risk-minimisation procedure of \cite{antoniadis09} have been proposed in the literature to select $h$ for functional kernel predictors. We adopt the former, in the specific form of a leave-future-out cross-validation, which is better suited to a forecasting context than a leave-one-out scheme, since it never allows the validation set to include curves observed after the point being predicted.

Concretely, let $n_\text{tr}$ denote the size of the training sample and let
$\mathcal{V} = \{n_\text{tr}-r+1,\, n_\text{tr}-r+2,\, \dots,\, n_\text{tr}\}$
denote the $r$ most recent days of the training sample used for validation.
For each $i\in\mathcal{V}$, the one-step-ahead prediction of $\mathcal{X}_{i+1}$
is formed using only the curves $\{\mathcal{X}_1,\dots,\mathcal{X}_{i-1}\}$
observed strictly before day $i$,
\begin{equation}\label{loo-pred}
\widehat{\mu}_h^{(-i)}(\mathcal{X}_i)
:= \sum_{j=1}^{i-1} \mathcal{W}_h(\mathcal{X}_i, \mathcal{X}_j)\,\mathcal{X}_{j+1},
\qquad
\mathcal{W}_h(\mathcal{X}_i, \mathcal{X}_j)
= \frac{\mathcal{K}\!\left(\dfrac{d(\mathcal{X}_i,\mathcal{X}_j)}{h}\right)}
       {\displaystyle\sum_{k=1}^{i-1}\mathcal{K}\!\left(\dfrac{d(\mathcal{X}_i,\mathcal{X}_k)}{h}\right)}.
\end{equation}
The leave-future-out cross-validation score for the mean is then
\begin{equation}\label{CV-mu}
\texttt{CV}_\mu(h)
= \frac{1}{r}\sum_{i\in\mathcal{V}}
  \left\|\mathcal{X}_{i+1} - \widehat{\mu}_h^{(-i)}(\mathcal{X}_i)\right\|^2,
\end{equation}
where $\|\cdot\|$ is the $L^2([0,\tau))$ norm, approximated in practice as
$\frac{1}{p}\sum_{j=1}^{p}\big(\mathcal{X}_{i+1}(t_j)
- \widehat{\mu}_h^{(-i)}(\mathcal{X}_i)(t_j)\big)^2$.
The optimal mean bandwidth is $h_\mu^* = \operatorname{argmin}_{h\in\mathcal{H}}\,\texttt{CV}_\mu(h)$,
where $\mathcal{H} = \{h_1,\dots,h_K\}$ is a finite candidate grid.

The identical procedure is applied a second time, independently, to select the variance bandwidth $h_\sigma$. The response series is replaced by the squared-residual curves $\widehat{R}_i$ defined in Eq.~\eqref{estU}, yielding the leave-future-out score
\begin{equation}\label{CV-sigma}
\texttt{CV}_\sigma(h)
= \frac{1}{r}\sum_{i\in\mathcal{V}}
  \left\|\widehat{R}_{i+1} - \widehat{\sigma}_h^{2,(-i)}(\mathcal{X}_i)\right\|^2,
\end{equation}
and the optimal variance bandwidth is $h_\sigma^* = \operatorname{argmin}_{h\in\mathcal{H}}\,\texttt{CV}_\sigma(h)$.
Because the conditional variance is typically a less smooth, more locally variable function of the conditioning curve than the conditional mean, the two bandwidths need not, and in general do not, coincide. For the national-level application of Section~\ref{sec4}, a grid of ten candidate bandwidths spanning $[2{,}000, 20{,}000]$~MW (in the units of the wavelet-coefficient semi-metric) was searched with a validation window of $r=60$ of the most recent training days; this procedure selected $h_\mu = 2{,}000$ for the conditional mean and $h_\sigma = 4{,}000$ for the conditional variance, confirming that the volatility surface required a wider neighbourhood --- and hence more averaging --- than the mean surface to be estimated reliably.

\subsubsection{Pointwise and uniform prediction bands\\}

\noindent\textbf{Heteroscedastic curve prediction.} 
Given a conditioning curve $\chi^\star\in \mathcal{E}$ (e.g., $\chi^\star=\mathcal X_n$), we estimate $\mu(\cdot)$ and $\sigma(\cdot)$ operators nonparametrically as described in Equations \eqref{NWE} and \eqref{estU}. Throughout this subsection, to avoid division by a near-zero estimated volatility, $\widehat\sigma_h(\cdot)$ denotes the floored quantity $\max\{\widehat\sigma_h(\cdot),\varepsilon\}$ for some fixed $\varepsilon>0$; this floor is omitted from subsequent displays for readability. Then, a prediction of the curve $\mathcal{X}_{n+1}$ is given as follows:
\begin{equation}
    \label{simuX}
    \widehat{\mathcal{X}}_{n+1}(t) = \widehat{\mu}_h(\chi^\star)(t) + \widehat{\sigma}_h(\chi^\star)(t) \times \zeta_{n+1}(t), \quad t\in [0, \tau),
\end{equation}
where $\zeta\stackrel{\text{i.i.d.}}{\sim}\mathbb P_\zeta,$
for a specified innovation law $\mathbb P_\zeta$ with $\mathbb E[\zeta(t)]=0.$

In practice, one can for instance simulate the paths of $\zeta$ as follows
\begin{equation}\label{zeta}
\zeta_i(t) = B_i(t) + N_i/ \sqrt{1-t(1-t)},\qquad i = 1, \dots, n,
\end{equation}
where $B_i(t)$ are i.i.d. standard Brownian Bridges and $N_i$ are i.i.d. standard normals.\\

\noindent\textbf{Prediction bands.}
In this section we introduce two methods to build predictive bands. \vskip 3mm
\noindent {\bf Method 1: Pointwise (marginal) Prediction Bands (PPB)}\\
For each fixed time $t\in[0,\tau)$, let
$
F_t(x|\chi^\star)
=
\mathbb{P}\left(\mathcal X_{n+1}(t)\le x | \chi^\star\right)
$
denote the conditional distribution function, and define the corresponding lower and upper quantiles
\begin{equation*}
q_\alpha(t|\chi^\star)=\inf\{x:\ F_t(x|\chi^\star)\ge \alpha\},
\qquad
q_{1-\alpha}(t|\chi^\star)=\inf\{x:\ F_t(x|\chi^\star)\ge 1-\alpha\}.
\end{equation*}
A $(1-2\alpha)$ pointwise predictive band at time $t$ is the interval $[q_\alpha(t| \chi^\star),\,q_{1-\alpha}(t| \chi^\star)]$.

Since $F_t(\cdot|\chi^\star)$ is unknown, we approximate it by Monte Carlo simulations. For this, generate $S$ next-day curves
$\{\widehat{\mathcal X}^{(s)}_{n+1}\}_{s=1}^S$ conditionally on $\chi^\star$ by using either:\\
(i) \emph{parametric} innovation generator
$\zeta^{(s)}$ given in Equation \eqref{zeta}, to obtain
\begin{equation}
\widehat{\mathcal X}^{(s)}_{n+1}(t)
=
\widehat\mu_h(\chi^\star)(t)
+
\widehat\sigma_h(\chi^\star)(t)\,\zeta^{(s)}_{n+1}(t).
\label{eq:param-scenarios}
\end{equation}
or \\
(ii) a \emph{bootstrap} generator based on standardized residual curves
\begin{equation}
\widehat Z_i(t)
=
\frac{\mathcal X_{i+1}(t)-\widehat\mu_h(\mathcal X_i)(t)}
{\widehat\sigma_h(\mathcal X_i)(t)},
\qquad i=1,\dots,n-1.
\label{eq:Zi-def}
\end{equation}
The standardized residuals $\widehat Z_i$ are resampled i.i.d.\ to form $Z^{(s)}$, and then mapped as
\begin{equation}
\widehat{\mathcal X}^{(s)}_{n+1}(t)
=
\widehat\mu_h(\chi^\star)(t)
+
\widehat\sigma_h(\chi^\star)(t)\,Z^{(s)}(t).
\label{eq:boot-scenarios}
\end{equation}
For each grid point $\{t_j\}_{j=1}^m$, define the empirical marginal distribution
\begin{equation}
\widehat F_{t_j}(x|\chi^\star)
=
\frac{1}{S}\sum_{s=1}^S \mathbf \1\!\left\{\widehat{\mathcal X}^{(s)}_{n+1}(t_j)\le x\right\},
\end{equation}
where $\1\{A\}$ is the indicator function that takes 1 if $A$ occurs and zero otherwise, and $\widehat{\mathcal X}_{n+1}^{(s)}(t)$ is generated either according to the curve prediction process given in \eqref{eq:param-scenarios} to obtain a Parametric Pointwise Prediction Band (PPPB) or \eqref{eq:boot-scenarios} to get a Bootstrap Pointwise Prediction Band (BPPB).

Then, one obtains the empirical quantiles
\begin{equation}
\widehat q_\alpha(t_j|\chi^\star) = \inf\{x:\ \widehat F_{t_j}(x|\chi^\star)\ge \alpha\},
\quad\text{and}\quad
\widehat q_{1-\alpha}(t_j|\chi^\star) = \inf\{x:\ \widehat F_{t_j}(x|\chi^\star)\ge 1-\alpha\}.
\end{equation}
The pointwise prediction band function on the grid is
\begin{equation}
t_j \ \longmapsto\ \big[\widehat q_\alpha(t_j|\chi^\star),\ \widehat q_{1-\alpha}(t_j|\chi^\star)\big],
\qquad j=1,\ldots,m,
\end{equation}
which can be interpolated to a continuous band on $[0,\tau).$

\begin{Remark}
Under the consistency of $\widehat\mu_h(\chi^\star)$ and $\widehat\sigma_h(\chi^\star)$, a correctly specified parametric generator for $\zeta$, and $S\to\infty$---the empirical quantiles satisfy
$\widehat q_\alpha(t_j|\chi^\star)\to q_\alpha(t_j|\chi^\star)$ and $\widehat q_{1-\alpha}(t_j| \chi^\star)\to q_{1-\alpha}(t_j| \chi^\star)$ for each fixed $t_j$. Consequently, for each $t_j$,
\begin{equation}
\mathbb{P}\!\left\{\mathcal X_{n+1}(t_j)\in\big[\widehat q_\alpha(t_j| \chi^\star),\,\widehat q_{1-\alpha}(t_j| \chi^\star)\big]\right\}
\ \longrightarrow\ 1-2\alpha.
\end{equation}
This coverage is \emph{marginal in time} and it holds pointwise for each $t$, but does not imply uniform (curve-wise) coverage over $t\in[0,\tau)$. In practice, $S$ should be large enough to stabilize tail quantiles of $\widehat F_{t_j}.$\\
We state these conditions to motivate the construction informally; verifying them rigorously for the present estimator is beyond the scope of this paper and is left for future work (see Section~\ref{sec6}).
\end{Remark}
%\vskip 3mm
\noindent {\bf Method 2: Uniform Prediction Bands (UPB)}\\
A $(1-\alpha)$ simultaneous (uniform-in-time) prediction band for $\mathcal X_{n+1}$ is a random set $\mathcal B_{\text{sim},\alpha}\subset \mathcal{E}$
such that $\mathbb{P}\!\big\{ \mathcal X_{n+1}(\cdot) \in \mathcal B_{\text{sim},\alpha} \,\big|\, \chi^\star \big\}
\approx 1-\alpha,$
i.e., the realized curve lies entirely inside the band with probability approximately $1-\alpha$.

To build this band we construct $S$ next-day scenarios conditionally on $\chi^\star$ by either:
\begin{itemize}
\item[(i)] \emph{parametric} innovation generator
$\zeta^{(s)}$, to obtain
\begin{equation}
\widehat{\mathcal X}^{(s)}_{n+1}(t)
=
\widehat\mu_h(\chi^\star)(t)
+
\widehat\sigma_h(\chi^\star)(t)\,\zeta^{(s)}_{n+1}(t),
\qquad \zeta^{(s)}\stackrel{\text{i.i.d.}}{\sim}\mathbb P_\zeta,
\label{eq:sim-param_2}
\end{equation}
\noindent or
\item[(ii)] a \emph{bootstrap} generator based on the standardized residual curves $\widehat Z_i(t)$ defined in Eq.~\eqref{eq:Zi-def} of Method 1 (computed, as there, by leave-one-out), resampled i.i.d.\ to form $Z^{(s)} \in \{\widehat Z_1,\ldots,\widehat Z_{n-1}\}$, and mapped as
\begin{equation}
\widehat{\mathcal X}^{(s)}_{n+1}(t)
=
\widehat\mu_h(\chi^\star)(t)
+
\widehat\sigma_h(\chi^\star)(t)\,Z^{(s)}(t).
\label{eq:sim-boot_2}
\end{equation}
\end{itemize}

For each scenario (parametric or bootstrap), standardize on the prediction scale:
\begin{equation*}
Z^{(s)}_{\text{std}}(t)
=
\frac{\widehat{\mathcal X}^{(s)}_{n+1}(t)-\widehat\mu_h(\chi^\star)(t)}
{\widehat\sigma_h(\chi^\star)(t)},
\qquad s=1,\ldots,S.
\end{equation*}
Note that, since $\widehat{\mathcal X}^{(s)}_{n+1}$ is itself constructed from $\widehat\mu_h(\chi^\star)$ and $\widehat\sigma_h(\chi^\star)$ via Eq.~\eqref{eq:sim-boot_2} (or its parametric counterpart \eqref{eq:sim-param_2}), this standardization step recovers $Z^{(s)}_{\text{std}}(t) \equiv Z^{(s)}(t)$ (resp.\ $\zeta^{(s)}(t)$) exactly; we retain it in the general form above since it remains valid under any scenario generator, including ones not tied to the same $\widehat\sigma_h(\chi^\star)$ used here.

Compute the discrete sup-norm
\begin{equation*}
M^{(s)}=\max_{1\le j\le m}\,\big|Z^{(s)}_{\text{std}}(t_j)\big|.
\end{equation*}
Let $\widehat c_\alpha
=
\mathrm{Quantile}_{1-\alpha}\big(\{M^{(s)}\}_{s=1}^S\big)
$
be the empirical $(1-\alpha)$ quantile of the suprema. The \emph{uniform prediction band} for $\mathcal{X}_{n+1}$ is
\begin{equation}
\ \mathcal B_{\text{sim},\alpha}
=
\Big\{ f\in \mathcal{E}:\ \max_{1\le j\le m}
\Big|
\dfrac{f(t_j)-\widehat\mu_h(\chi^\star)(t_j)}{\widehat\sigma_h(\chi^\star)(t_j)}
\Big|
\le \widehat c_\alpha
\Big\}.
\label{eq:sim-band}
\end{equation}
Equivalently, for all time points $t_1, \dots t_m$ on the grid, one has
\begin{equation}
\mathcal{X}_{n+1}(t_j)\in
\Big[
\widehat\mu_h(\chi^\star)(t_j)-\widehat c_\alpha\,\widehat\sigma_h(\chi^\star)(t_j),
\ \widehat\mu_h(\chi^\star)(t_j)+\widehat c_\alpha\,\widehat\sigma_h(\chi^\star)(t_j)
\Big],
\qquad \forall j\in \{1,\ldots,m\}.
\end{equation}

\begin{Remark}
Under uniform consistency of $\widehat\mu_h(\chi^\star)$ and $\widehat\sigma_h(\chi^\star)$, a dense grid
($\max_j(t_{j+1}-t_j)\to 0$), and either (i) correct parametric specification in \eqref{eq:sim-param_2} so that
$\mathcal L\left(\sup_{t\in[0,\tau]}|Z_{n+1}^\circ(t)|\right)=\mathcal L(M^{(s)})$ where
$Z_{n+1}^\circ(t)=\{\mathcal X_{n+1}(t)-\mu(\chi^\star)(t)\}/\sigma(\chi^\star)(t)$, and $\mathcal{L}$ is its probability distribution, or (ii) bootstrap consistency in \eqref{eq:sim-boot_2}, the empirical quantile
$\widehat c_\alpha$ consistently estimates the population threshold $c_\alpha$ of $\sup_{t\in[0,\tau)}|Z_{n+1}^\circ(t)|$. Consequently,
\begin{equation}
\mathbb{P}\!\big\{ \mathcal X_{n+1} \in \mathcal B_{\text{sim},\alpha} \big\}
\ \longrightarrow\ 1-\alpha
\qquad \text{as } S\to\infty,\ m\to\infty.
\end{equation}
We state these conditions to motivate the construction informally; verifying them rigorously for the present estimator is beyond the scope of this paper and is left for future work (see Section~\ref{sec6}).
\end{Remark}

\section{Accuracy measures}\label{sec:accuracy}

A band can always be made arbitrarily narrow at the cost of failing to contain the realised curve, or arbitrarily wide at the cost of being uninformative, and neither failure is visible unless width and validity are both assessed \citep{gneiting14}. Following the evaluation protocol used for functional-data prediction intervals in \cite{antoniadis2016}, we report, for a band $\big[L_{j,\alpha}(t_h), U_{j,\alpha}(t_h)\big]$ constructed at nominal level $1-\alpha$ for day $j$ and half-hour $t_h$, $h=1,\dots,48$, over a test sample of $n$ days, the following three criteria.

\subsection*{Mean amplitude.} The average width of the band,
$$
\texttt{Amp}_\alpha = \frac{1}{n}\sum_{j=1}^n \frac{1}{48}\sum_{h=1}^{48} \Big(U_{j,\alpha}(t_h) - L_{j,\alpha}(t_h)\Big),
$$
measures the \emph{sharpness} of the band: how informative it is, independently of whether it is correct. A narrower band is preferable, all else being equal, but amplitude is only meaningful when read jointly with coverage, since width can trivially be reduced by under-stating uncertainty.

\subsection*{Mean pointwise coverage.} The empirical frequency, across the test sample, with which the realised curve falls inside the band at a given half-hour, averaged over the 48 half-hours of the day,
$$
\texttt{Cov}_\alpha = \frac{1}{n}\sum_{j=1}^n \frac{1}{48}\sum_{h=1}^{48} \1\big\{\mathcal{X}_j(t_h) \in \big[L_{j,\alpha}(t_h), U_{j,\alpha}(t_h)\big]\big\},
$$
measures \emph{calibration}: whether a band advertised at nominal level $1-\alpha$ in fact contains the truth with that frequency. A method whose $\texttt{Cov}_\alpha$ falls systematically below $1-\alpha$ is overconfident, irrespective of how narrow or wide its bands appear in isolation.

\subsection*{Curve-wise coverage.} Pointwise coverage is a marginal criterion: it can be satisfied on average across the 48 half-hours of a day even when the realised curve, taken as a whole path rather than as 48 separate scalars, is essentially never contained entirely within the band. Since a load curve is used operationally as a single trajectory --- for scheduling, reserve sizing, or risk management over the full day-ahead horizon --- the relevant notion of coverage for such decisions is simultaneous rather than marginal. We therefore also report the curve-wise coverage,
$$
\texttt{CW}_{\alpha,k} = \frac{1}{n}\sum_{j=1}^n \1\Bigg\{ \sum_{h=1}^{48} \1\big\{\mathcal{X}_j(t_h) \notin \big[L_{j,\alpha}(t_h), U_{j,\alpha}(t_h)\big]\big\} \le k \Bigg\},
$$
the proportion of test days on which at most $k$ of the 48 half-hourly points fall outside the band, for $k=0,1,2,3$, following the construction proposed by \cite{antoniadis2016}. Setting $k=0$ recovers the strict requirement that the entire daily path be contained in the band; allowing $k>0$ accommodates a forecaster willing to tolerate a small, pre-specified number of localised exceptions while still requiring the bulk of the day to be correctly bounded. Reporting $\texttt{CW}_{\alpha,k}$ alongside $\texttt{Cov}_\alpha$ distinguishes a band that is valid only marginally, half-hour by half-hour, from one that delivers the stronger, operationally more meaningful guarantee of covering the whole forecasted trajectory.

Together, $\texttt{Amp}_\alpha$, $\texttt{Cov}_\alpha$ and $\texttt{CW}_{\alpha,k}$ give a complete sharpness--calibration picture of a prediction band at a chosen nominal level $1-\alpha$: the first measures how informative the band is, the second whether it is marginally valid, and the third whether it is valid at the level of the whole forecasted curve. All three are reported jointly, and at several nominal levels, in Section~\ref{sec4}, since no single one of them is sufficient on its own to judge whether a method's stated uncertainty can be trusted.

\section{Application to French national level electrical load curve}\label{sec4}

\begin{figure}[h]
\begin{center}
\includegraphics[width=\textwidth]{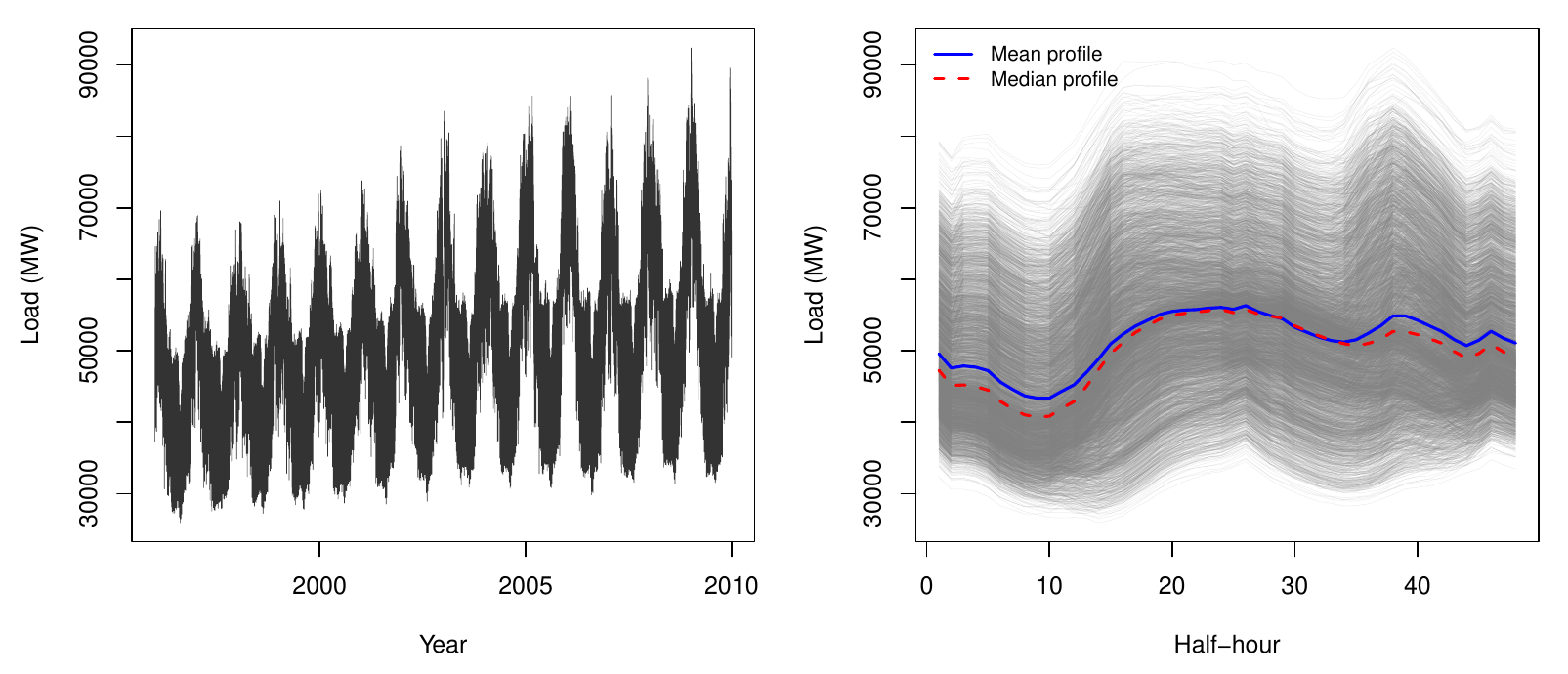}
\end{center}
\caption{(a) Half-hourly French national electricity demand, 01/01/1996--31/12/2009. (b) All 5{,}114 daily load curves, with the mean (solid) and median (dashed) daily profile.}
\label{load_nat}
\end{figure}

A half-hourly French national electricity consumption data set observed between 01/01/1996 and 31/12/2009 is considered. We focus on day-ahead forecast of the load curve for the year 2009. The model is fitted and the bandwidths $h_\mu$ and $h_\sigma$ are selected by leave-future-out cross-validation using data from 01/01/1996 to 31/12/2008, while the data from 01/01/2009 to 31/12/2009 is reserved for testing. Following \cite{antoniadis2016}, the 30 calendar days surrounding public holidays are excluded from the test set, leaving 332 test days. Figure~\ref{load_nat} displays the raw half-hourly demand series for the full sample period together with the resulting collection of 5{,}114 daily load curves; the marked annual cycle, the upward trend driven by demand growth, and the pronounced difference between the morning ramp/evening peak shape of working days and the flatter weekend profile are all visible in the overlaid curves.

The conditional mean $\widehat\mu_h(\cdot)$ is estimated as in Eq.~(\ref{NWE}), and the conditional variance $\widehat\sigma_h^2(\cdot)$ as in Eq.~(\ref{estU}), with the wavelet-based semi-metric and the leave-future-out bandwidths $h_\mu$, $h_\sigma$ described in Section~\ref{22}. Pointwise (\texttt{PPB}) and uniform, sup-norm (\texttt{UPB}) prediction bands are constructed from $S=1{,}000$ scenarios drawn around these estimates, as described in Section~\ref{22}. We use the bootstrap generator rather than the parametric one of Eq.~(\ref{zeta}) for two reasons. First, the parametric generator presupposes a specific innovation law $\mathbb{P}_\zeta$, built from a Brownian bridge and an independent normal term, and nothing in the data guarantees that the standardised innovations $\zeta(t)$ of national-level demand actually follow this, or any other, parametric law. Second, this matters empirically. Re-estimating both \texttt{FARCH} bands with the parametric generator in place of the bootstrap one, holding $\widehat\mu_h(\cdot)$ and $\widehat\sigma_h(\cdot)$ fixed, shows that the parametric pointwise band reaches only 55.5\% observed coverage at the 95\% nominal level, against 85.7\% for the bootstrap pointwise band, and that the parametric uniform band reaches 70.9\%, against 96.8\% for the bootstrap uniform band --- a calibration shortfall far larger than any difference reported between methods in Tables~\ref{table2_coverage}--\ref{table3_curvewise}. The bootstrap construction, by contrast, imposes no parametric assumption on $\zeta(t)$ and instead preserves the empirical shape of the observed residual curves.

We compare these two \texttt{FARCH}-based bands against the four prediction-interval constructions proposed by \cite{antoniadis2016} for their Kernel+Wavelet+Functional (\texttt{KWF}) predictor, which --- unlike \texttt{FARCH} --- does not estimate a conditional variance function. Instead, the dispersion of its prediction interval is inherited entirely from the spread of $B=500$ bootstrap pseudo-curves resampled with probability equal to the kernel weights themselves. We re-implemented all four variants on the present data set, using the same kernel weights and the same calendar-transition restriction as for \texttt{FARCH}, so that the comparison that follows isolates the effect of modelling $\sigma^2_h(\cdot)$ explicitly. The four variants are the symmetric Gaussian-quantile band (\texttt{S-KWF}), the asymmetric empirical-quantile band of the disconnected detail/approximation residuals (\texttt{NS-KWF}), the family-wise-error-controlled band that tolerates $k=2$ uncovered half-hours by construction (\texttt{$k$-FWE}, \citealp{wolf2013}), and the Nearest Path heuristic that iteratively discards the most extreme bootstrap trajectory (\texttt{NP}, \citealp{staszewska2011}).

Following \cite{antoniadis2016}, we report $\texttt{Amp}_\alpha$ and $\texttt{Cov}_\alpha$ (Section~\ref{sec:accuracy}) at the 80\%, 90\% and 95\% nominal levels (Tables~\ref{table1_amplitude}--\ref{table2_coverage}), and $\texttt{CW}_{\alpha,k}$ at the 95\% level for $k=0,1,2,3$ (Table~\ref{table3_curvewise}), exactly as in \citeauthor{antoniadis2016}'s own Table~3.

\begin{table}[h]
\renewcommand{\arraystretch}{1.2}
\caption{Mean amplitude $\texttt{Amp}_\alpha$ (MW) of the prediction band, by method and nominal confidence level, test year 2009.}
\label{table1_amplitude}
\centering
\begin{tabular}{lccc}
\toprule
Method & 80\% & 90\% & 95\% \\
\midrule
\texttt{FARCH-PPB} & 5{,}926  & 9{,}147  & 12{,}290 \\
\texttt{FARCH-UPB} & 13{,}001 & 17{,}328 & 21{,}799 \\
\texttt{$k$-FWE}   & 7{,}525  & 10{,}077 & 12{,}078 \\
\texttt{NP}        & 5{,}689  & 7{,}702  & 9{,}059  \\
\texttt{S-KWF}     & 4{,}798  & 7{,}306  & 9{,}378  \\
\texttt{NS-KWF}    & 4{,}218  & 6{,}482  & 8{,}091  \\
\bottomrule
\end{tabular}
\end{table}

\begin{table}[h]
\renewcommand{\arraystretch}{1.2}
\caption{Mean pointwise coverage $\texttt{Cov}_\alpha$ (\%) of the prediction band, by method and nominal confidence level, test year 2009.}
\label{table2_coverage}
\centering
\begin{tabular}{lccc}
\toprule
Method & 80\% & 90\% & 95\% \\
\midrule
\texttt{FARCH-PPB} & 55.7 & 74.0 & 85.7 \\
\texttt{FARCH-UPB} & \textbf{86.4} & \textbf{93.9} & \textbf{97.2} \\
\texttt{$k$-FWE}   & 66.6 & 76.9 & 82.5 \\
\texttt{NP}        & 55.4 & 66.4 & 71.2 \\
\texttt{S-KWF}     & 50.1 & 64.7 & 74.0 \\
\texttt{NS-KWF}    & 44.3 & 59.9 & 66.6 \\
\bottomrule
\end{tabular}
\end{table}

\begin{table}[h]
\renewcommand{\arraystretch}{1.2}
\caption{Curve-wise coverage $\texttt{CW}_{\alpha,k} $ (\%) at the 95\% nominal level: proportion of test days with at most $k$ half-hourly points outside the band.}
\label{table3_curvewise}
\centering
\begin{tabular}{lcccc}
\toprule
Method & $k=0$ & $k=1$ & $k=2$ & $k=3$ \\
\midrule
\texttt{FARCH-PPB} & 68.1 & 71.7 & 73.5 & 74.1 \\
\texttt{FARCH-UPB} & \textbf{90.1} & \textbf{92.2} & \textbf{94.0} & \textbf{94.6} \\
\texttt{$k$-FWE}   & 70.8 & 71.7 & 72.9 & 74.4 \\
\texttt{NP}        & 53.0 & 56.3 & 58.1 & 59.6 \\
\texttt{S-KWF}     & 58.1 & 58.7 & 59.6 & 61.4 \\
\texttt{NS-KWF}    & 47.3 & 49.1 & 51.2 & 54.2 \\
\bottomrule
\end{tabular}
\end{table}

Two patterns emerge clearly from Tables~\ref{table1_amplitude}--\ref{table3_curvewise}, and both replicate, on an independent national-level data set, the central empirical finding of \cite{antoniadis2016}. First, \emph{pointwise} band constructions that do not explicitly account for the dependence between half-hourly points within a day systematically undercover relative to their nominal level. \texttt{S-KWF} and \texttt{NS-KWF} reach only 74.0\% and 66.6\% observed coverage, respectively, against a 95\% target, mirroring the 88\% and 82\% obtained by the corresponding variants in \citeauthor{antoniadis2016}'s own French data. Tellingly, \texttt{FARCH-PPB} --- which does model $\sigma_h^2(\cdot)$ explicitly, but still constructs its band from marginal, time-by-time quantiles --- undercovers in essentially the same way (85.7\% at the 95\% level). This indicates that the pointwise-coverage shortfall is a structural property of marginal band construction itself, not an artefact of ignoring conditional heteroscedasticity, since even a correctly specified, state-dependent dispersion does not, on its own, deliver a band that is valid simultaneously across the 48 points of the curve.

\begin{figure}[h]
\begin{center}
\includegraphics[width=0.85\textwidth]{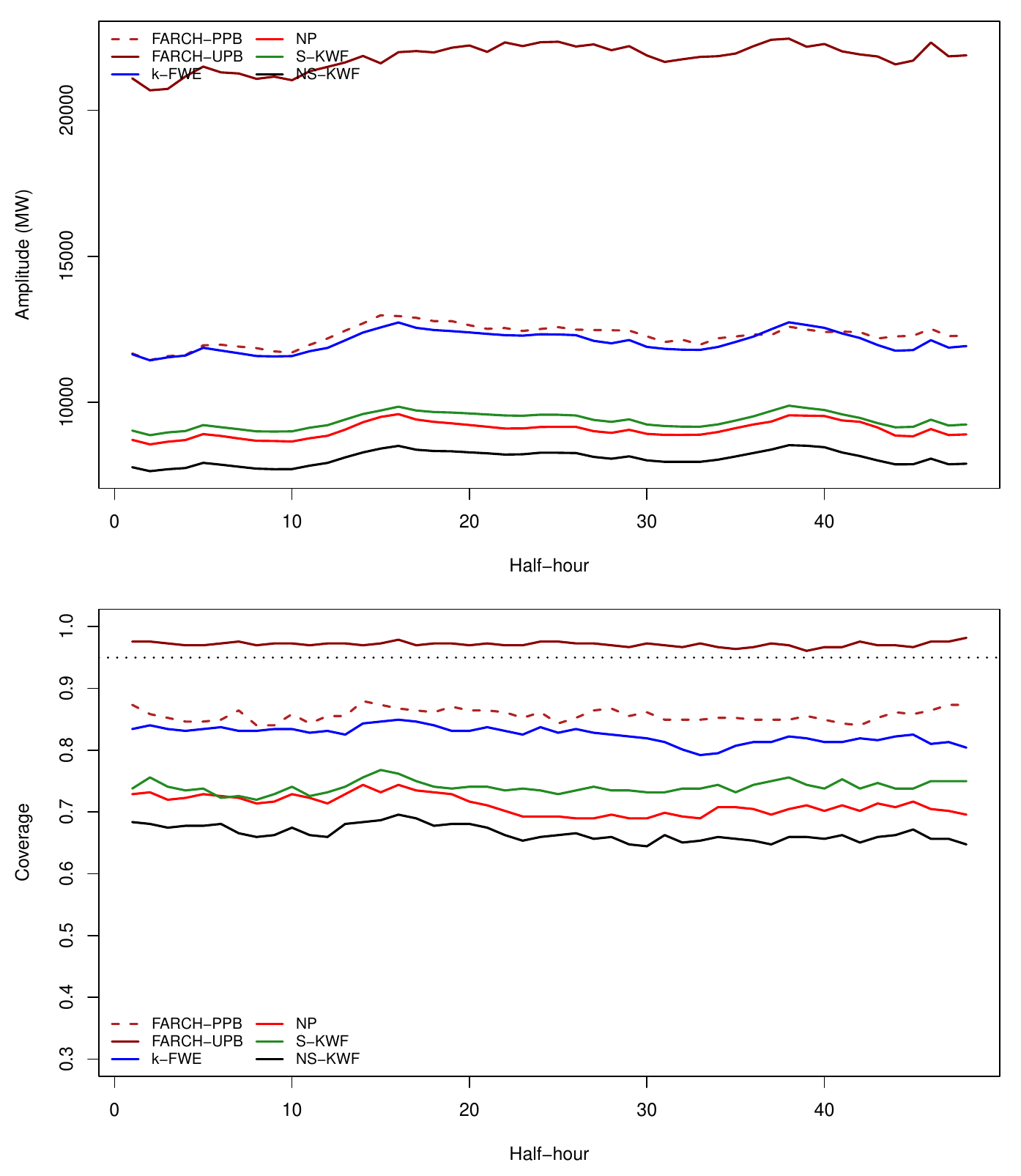}
\end{center}
\caption{Mean amplitude (top) and mean coverage (bottom) by half-hour, 95\% confidence level.}
\label{fig_byhour}
\end{figure}

Figure~\ref{fig_byhour} disaggregates both the amplitude and the coverage of all six methods by half-hour, at the 95\% level, and shows exactly where this shortfall is concentrated. The amplitude of every method tracks the daily demand cycle, widening during the morning ramp and the late-afternoon peak and narrowing overnight when demand --- and its variability --- is lowest, consistent with the heteroscedastic, time-of-day-dependent scale function that motivates the \texttt{FARCH} construction in the first place. The coverage panel shows that every pointwise or bootstrap-only method's coverage degrades, to varying degrees, precisely during the high-demand, high-volatility hours in the middle of the day --- the hours at which a forecasting error is operationally most costly --- while \texttt{FARCH-UPB} remains within a narrow band around its 95\% target throughout the day.

\begin{figure}[h]
\begin{center}
\includegraphics[width=\textwidth]{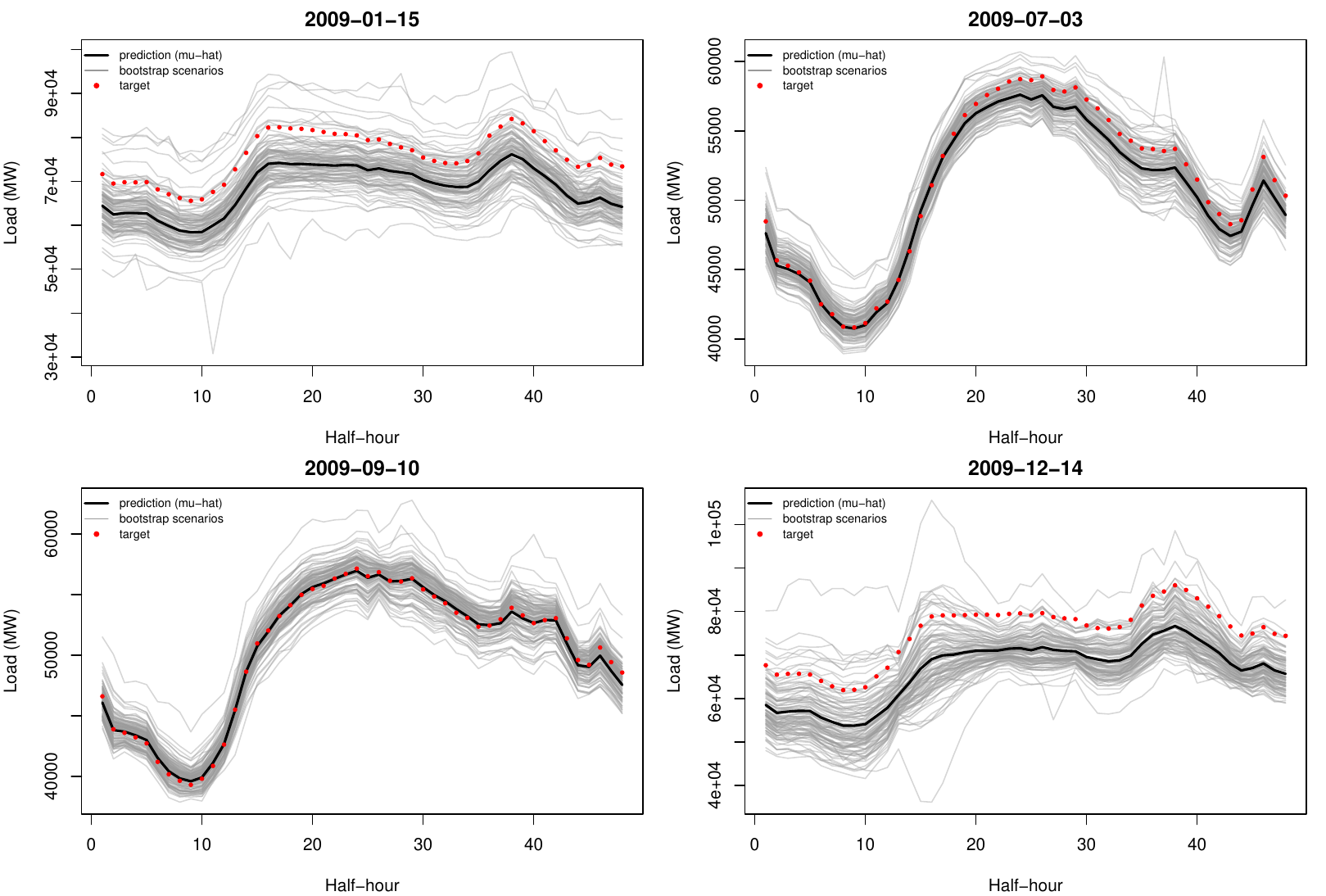}
\end{center}
\caption{\texttt{FARCH} bootstrap scenario fan (grey), point forecast (black) and realised curve (red dots) for four days of the 2009 test year.}
\label{fig_selected_cases}
\end{figure}

Figure~\ref{fig_selected_cases} illustrates the mechanism underlying this pattern at the level of individual days, using \texttt{FARCH}'s own bootstrap scenario fan, point forecast and realised curve for four selected days: two days with stable, easily forecastable demand (3 July and 10 September 2009) and two days falling in the more volatile winter regime (15 January and 14 December 2009). On the two summer days the scenario fan is narrow and the target sits comfortably within it throughout the day; on the two winter days the fan widens substantially, reflecting the larger estimated $\widehat\sigma_h(\cdot)$, and the realised curve occasionally tests the edge of the fan around the evening peak --- precisely the behaviour a state-dependent volatility function is designed to produce, and the same qualitative pattern documented by \cite{antoniadis2016} for their own selected cases.

\begin{figure}[h]
\begin{center}
\includegraphics[width=0.75\textwidth]{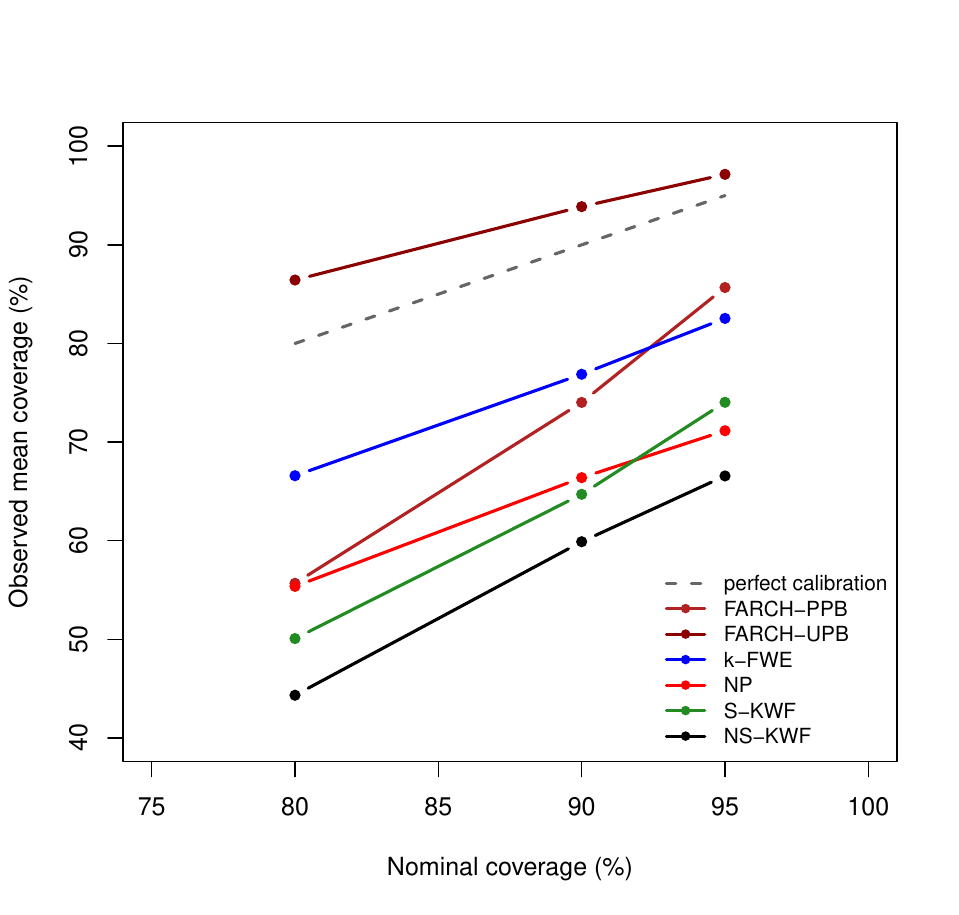}
\end{center}
\caption{Observed versus nominal coverage, all six methods, at the 80\%, 90\% and 95\% confidence levels. The dashed diagonal is the locus of perfect calibration.}
\label{fig_calibration}
\end{figure}

Second, and in direct response to the pointwise shortfall documented above, the bands that are built to control the \emph{simultaneous} (curve-wise) coverage of the whole path --- \texttt{$k$-FWE} on one side, and our uniform sup-norm construction \texttt{FARCH-UPB} on the other --- both improve markedly on the pointwise variants, but only \texttt{FARCH-UPB} achieves coverage close to its nominal target at every level considered (86.4/93.9/97.2\% observed against 80/90/95\% nominal, Table~\ref{table2_coverage}), and a curve-wise coverage of 90.1\% with zero half-hours allowed to fall outside the band (see Table~\ref{table3_curvewise}). By comparison, \texttt{$k$-FWE} --- despite being explicitly calibrated to tolerate $k=2$ uncovered half-hours, a strictly more permissive criterion --- reaches only 70.8\% curve-wise coverage at $k=0$ and 82.5\% pointwise coverage at the 95\% level on this data set. A paired comparison across the 332 test days confirms that this gap is not sampling noise, since both the amplitude and the coverage differences between \texttt{FARCH-UPB} and \texttt{$k$-FWE} are significant at $\text{p-value} <2.2\times10^{-16}$. Figure~\ref{fig_calibration} summarises this comparison in a single calibration diagram that plots observed against nominal coverage for all six methods. \texttt{FARCH-UPB} is the only curve that tracks the diagonal of perfect calibration across the full range, while every \texttt{KWF}-family variant, including \texttt{$k$-FWE}, lies visibly and consistently below it.

\begin{figure}[h]
\begin{center}
\includegraphics[width=0.7\textwidth]{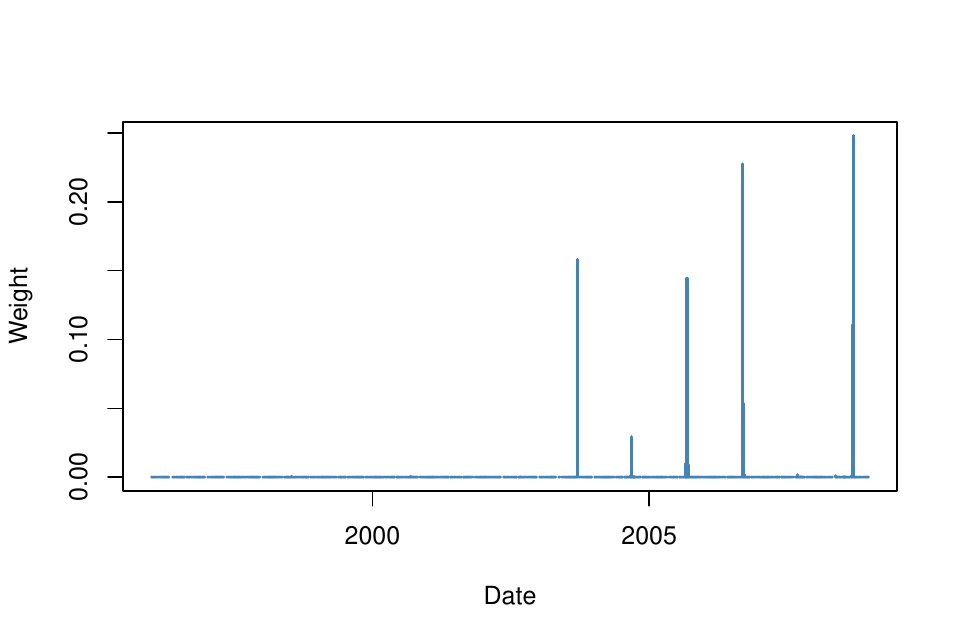}
\end{center}
\caption{\texttt{FARCH} kernel weight vector used to predict 10 September 2009, plotted against the calendar date of the contributing historical curve.}
\label{fig_weights}
\end{figure}

The mechanism behind this difference is, we believe, the explanation that \cite{antoniadis2016} themselves point to when motivating their own simultaneous-interval constructions. The \texttt{KWF} bootstrap pseudo-curves carry dispersion information only through the spread of resampled historical analogues, which is a function of how heterogeneous the matched historical pool happens to be, not of an explicitly estimated, state-dependent scale. \texttt{FARCH} ties the width of its uniform band directly to $\widehat{\sigma}_h(\chi)$, so that the band widens or narrows with the conditioning curve's own estimated volatility regime rather than with the incidental composition of its bootstrap neighbourhood. Figure~\ref{fig_weights} offers a brief check on this mechanism, reporting the weight vector used to predict 10 September 2009 --- the same calendar date used by \cite{antoniadis2016} for their own illustrative example (their Fig.~5, for the year 2006). As in their analysis, the weights are concentrated on historical occurrences of the same date in earlier years (2003, 2006 and 2008 receive the largest weights here), confirming that the wavelet semi-metric is extracting genuine seasonal structure rather than an artefact of the construction.

The cost of this better calibration is visible in Table~\ref{table1_amplitude}. \texttt{FARCH-UPB} produces the widest interval of all six methods at every confidence level (21{,}799~MW at the 95\% level, against 12{,}078~MW for \texttt{$k$-FWE}), or equivalently 39.5\% of the test year's mean load level of 55{,}248~MW. This is the expected price of a genuine simultaneous guarantee, and we report it for what it is: a calibrated band is, by definition, only as narrow as the data allow, and a narrower interval purchased by under-stating coverage is not a cheaper one, only a less honest one.

We stress that the comparison above is internal. Both the \texttt{FARCH} bands and all four \texttt{KWF}-family baselines were estimated and evaluated on the same data, the same train/test split, the same calendar-transition restriction and the same bootstrap protocol, which is what allows the paired significance tests reported above to be informative. We do not claim that the absolute figures in Tables~\ref{table1_amplitude}--\ref{table3_curvewise} improve on the absolute figures reported in \citeauthor{antoniadis2016}'s own Tables~1--2, since their data set is a distinct, proprietary EDF series, evaluated over a different test period and, to judge from the load scale plotted in their Fig.~1, aggregated at a different level than the present national series (mean load 55{,}248~MW here); a numerical comparison across the two studies would conflate the effect of the method with the effect of the data. What the present, controlled comparison does establish is that, on identical national-level French demand data and under an identical evaluation protocol, replacing the implicit, bootstrap-only dispersion of the \texttt{KWF} construction with an explicitly estimated conditional volatility function yields a uniform prediction band whose coverage is statistically indistinguishable from its nominal target, while every variant that foregoes such a model --- including the best of \citeauthor{antoniadis2016}'s own proposals, \texttt{$k$-FWE} --- falls significantly short of it on this data.

\section{Conclusion}\label{sec6}

This paper has introduced \texttt{FARCH}, a functional nonlinear autoregressive model with \texttt{ARCH}-type errors, in which the conditional mean and the conditional volatility of an intraday load curve are estimated jointly and nonparametrically, by reusing a single kernel-weighting scheme built on a wavelet semi-metric between curves, first on the curves themselves and then on their squared residuals. From this pair of nonparametric estimators, two constructions of prediction bands were derived, pointwise and uniform, the latter calibrated from the sup-norm of standardised scenario paths, intended respectively for local and for whole-path operational decisions.

Applied to day-ahead forecasting of the French national load, and compared under an identical protocol against four established constructions for the \texttt{KWF} predictor that do not model the conditional variance explicitly, the uniform \texttt{FARCH} band was the only method whose observed coverage tracked its nominal target across all three confidence levels considered, and the only one to deliver a curve-wise coverage close to its announced level, with both improvements found to be statistically significant. This better calibration was not free, since the uniform \texttt{FARCH} band was also the widest of the six methods compared, which we take to be the expected price of a genuine simultaneous guarantee rather than a shortcoming to be explained away. A further, more general finding emerged from this comparison, namely that pointwise band constructions undercovered relative to their nominal level whether or not the conditional variance was modelled explicitly. This suggests that marginal calibration and state-dependent volatility are two distinct properties, and that a load-curve prediction band needs both --- an explicit volatility function \emph{and} a simultaneous, path-level construction --- to be trusted as a description of forecast uncertainty rather than as a pointwise artefact.

Looking ahead, this work opens several natural directions. A first perspective is theoretical, namely establishing the consistency of $\widehat\mu_h$ and $\widehat\sigma_h$, building on results already available for related nonparametric functional estimators, which would place the calibration documented in this paper on the same firm theoretical footing as the kernel-weighting scheme itself. A second perspective is to enrich the conditioning information itself, since both $\mu(\cdot)$ and $\sigma(\cdot)$ are currently estimated from the shape of the previous load curve alone. Exogenous covariates such as a genuine day-ahead temperature or humidity forecast, rather than their realised values, could be incorporated directly into the volatility function, allowing the width of the prediction band to widen in anticipation of an unusually cold or unstable day rather than only after the fact. Such covariates could be combined with the wavelet semi-metric already used here, through an extended distance that weighs curve shape and forecast weather jointly, or introduced as a separate conditioning step applied to the same kernel-weighting scheme.

\end{document}